\documentclass[conference]{IEEEtran}
\IEEEoverridecommandlockouts
\usepackage{cite}
\usepackage{url}
\usepackage{array}
\usepackage{booktabs}
\usepackage{amsmath,amssymb,amsfonts}
\usepackage{algorithmic}
\usepackage{graphicx}
\usepackage{textcomp}
\usepackage{xcolor}
\usepackage{makecell} 
\usepackage{multirow}
\usepackage[margin=1in]{geometry}
\def\BibTeX{{\rm B\kern-.05em{\sc i\kern-.025em b}\kern-.08em
    T\kern-.1667em\lower.7ex\hbox{E}\kern-.125emX}}
\begin{document}

\title{RE-LLM: Refining Empathetic Speech-LLM Responses by Integrating Emotion Nuance}

\author{
\IEEEauthorblockN{Jing-Han Chen\IEEEauthorrefmark{1},
Bo-Hao Su\IEEEauthorrefmark{1},
Ya-Tse Wu\IEEEauthorrefmark{1},
Chi-Chun Lee\IEEEauthorrefmark{1}}

\IEEEauthorblockA{\IEEEauthorrefmark{1}
Department of Electrical Engineering, National Tsing Hua University\\
Hsinchu, Taiwan\\
Email: \{hannthu1915, borrissu, crowpeter\}@gapp.nthu.edu.tw, cclee@ee.nthu.edu.tw}
}

\maketitle

\begin{abstract}

With generative AI advancing, empathy in human-AI interaction is essential. While prior work focuses on emotional reflection, emotional exploration—key to deeper engagement—remains overlooked. Existing LLMs rely on text which captures limited emotion nuances. To address this, we propose RE-LLM, a speech-LLM integrating dimensional emotion embeddings and auxiliary learning. Experiments show statistically significant gains in empathy metrics almost across three datasets. RE-LLM relatively improves the Emotional Reaction score by 14.79\% and 6.76\% compared to text-only and speech-LLM baselines on ESD. Notably, it raises the Exploration score by 35.42\% and 3.91\% on IEMOCAP, 139.28\% and 9.83\% on ESD and 60.95\% and 22.64\% on MSP-PODCAST relatively. It also boosts unweighted accuracy by 5.4\% on IEMOCAP, 2.3\% on ESD and 6.9\% on MSP-PODCAST in speech emotion recognition. These results highlight the enriched emotional understanding and improved empathetic response generation of RE-LLM.

\end{abstract}


\section{Introduction}

As generative AI technology matures, interactions between humans and AI agents have become increasingly frequent. Research highlights the importance of \textit{empathy} in these interactions. Empathetic AI not only influences the dynamics of human-AI interactions but also significantly impacts the perceived quality \cite{liu2024illusionempathyaichatbots}.
By fostering a sense of understanding and connection, empathy enhances user satisfaction of conversations and trust \cite{rostami2023artificial}.
In terms of the attitude of human to AI agent, empathetic AI also gains more acceptance and more long-term user engagement \cite{bickmore2005establishing,pelau2021makes,leite2014empathic}, leading to strengthening loyalty and creating values \cite{liu2022artificial}. Furthermore, empathetic AI has the potential to transcend rigid frameworks, allowing for more personalized and tailored responses that adapt to individuality \cite{kidder2024empathyrightexceptionllms}.

Current empathetic LLMs are predominantly defined to generate \textit{empathetic} responses by responding with appropriate, often matching emotions \cite{chen2023soulchat,schaaff2023exploring,welivita2024largelanguagemodelsempathetic,lee2024large}. For instance, Yan \textit{et al.} \cite{yan2024talkhumanlikeagentsempathetic} train the empathetic speech-LLM to respond with reciprocal styles and tones. However, human's \textit{emphatic} responses are much more complex, studies indicate that exploratory questions are capable of eliciting deeper emotional expressions and feelings from participants that leads to a more empathetic engagement \cite{miller2003manual,elliott2011empathy,williams2023use} and makes the intervention to the client more effective and proper \cite{iwakabe2000relationship}. Furthermore, Rahmani \textit{et al.} \cite{rahmani2024clarifying} show that well-crafted clarifying questions improve information retrieval performance and user satisfaction, while neutral ones cause frustration and confusion. That is, besides simply matching emotional expressions, complex exploration capability is another critical aspect for advancing empathetic LLMs.

Furthermore, although various studies have worked on empathetic LLMs, there is a lack of systematic integration of emotion nuances manifested in speech signals. In fact, existing approaches capture emotion in paralinguistic speech cues through text caption that describes the speaking styles \cite{yan2024talkhumanlikeagentsempathetic} or ASR/transcription-based speech embeddings \cite{wang2024blspemoempatheticlargespeechlanguage}, which only reflects partial information, e.g., the verbal (content)-based emotion, overlooking affect-related nuances and subtleties in speech signal. We argue that this lack of emotion nuance modeling would lead to a suboptimal empathetic-LLM response generation, e.g., inability to adapt and respond with emotion exploration. In this work, we propose a combination of an emotion speech embedding and an emotion auxiliary task to enhance current speech-based empathetic LLMs without undergoing a series of complex fine-tuning of LLM. Specifically, our approach, terms as \textbf{RE-LLM}\footnote{\url{https://github.com/CHEN-JING-HAN/RE-LLM.git}}, incorporated richer speech emotion embeddings from wav2vec 2.0, a pretrained dimensional emotion prediction model \cite{wagner2023dawn}, and append them to speech embeddings as input for speech-LLM. We further introduce a dimensional emotion auxiliary task alongside the widely-used categorical one, enabling the LLM to better comprehend emotional information \cite{kang2024frozen}. 

We evaluate our framework using the two well-known empathetic metrics \cite{sharma2020empathy}: ``Emotional Reaction" score is a measurement of how properly the LLM responds to the emotion, which is also the most commonly referred metric, and another one is the ``Exploration" score that links to the ability to respond with an empathetic question.  Our results show that RE-LLM significantly enhances empathetic ability and that the improvements are statistically significant almost across three datasets, relatively improving the ``Emotional Reaction" score by 14.79\% and 6.76\% on ESD compared to the text-only LLM and the existing speech-LLM baseline, respectively. Likewise, RE-LLM also achieves notable improvements on MSP-PODCAST. For IEMOCAP, RE-LLM also shows a slight improvement in ``Emotional Reaction" scores. More importantly, RE-LLM achieves a significant improvement in the ``Exploration" score, having a relative increase of 35.42\% and 3.91\% on IEMOCAP, of 139.28\% and 9.83\% on ESD and of 60.95\% and 22.64\% on MSP-PODCAST compared to text-LLM and speech-LLM baseline, indicating a better and more complete empathetic capability. Additionally, the unweighted accuracy (UA) of speech emotion recognition improves by 5.4\% on IEMOCAP, 2.3\% on ESD and 6.9\% on MSP-PODCAST compared to pretrained speech-LLM baseline, demonstrating the model’s enhanced ability to perceive emotions accurately. This study highlights the importance of richer emotional embeddings and emotion-related auxiliary tasks in improving response generation from speech-based LLMs. By capturing nuanced emotional dynamics, our method shows a critical improvement in generating better empathetic responses.

\begin{figure*}[t]
\centering
  \includegraphics[width=14.21cm,height=5.7cm]{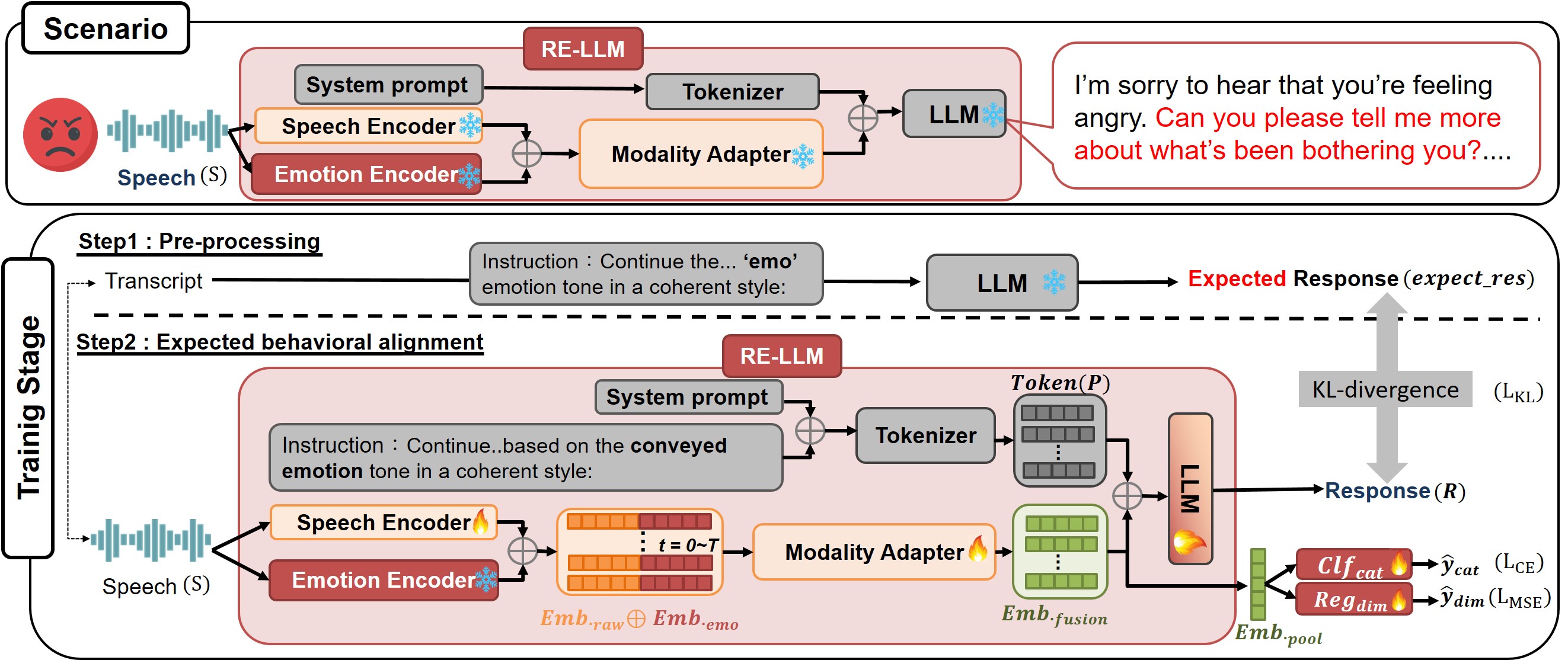}
  \caption{The architecture of our proposed RE-LLM comprises a speech-LLM and an emotion nuance module. A preprocessing generation and expected behavioral alignment constrained on nuance emotion training strategy are depicted as well.}
  \label{fig:framework}
  \vspace{-7mm}
\end{figure*}

\section{Research Methodology}
\subsection{RE-LLM}
We propose RE-LLM, a speech-LLM framework that enhances empathy in interactions. Inspired by the SOTA BLSP-Emo \cite{wang2024blspemoempatheticlargespeechlanguage}, we use it as our backbone. Like standard speech-LLMs, RE-LLM processes speech input via a speech encoder and a modality adapter, mapping speech embeddings into an LLM-understandable format for response generation. Contrary to previous speech-LLMs, RE-LLM improves empathetic responses by incorporating multi-perspective emotion nuances. Specifically, we develop an emotion nuance module in RE-LLM, which consists of an emotion encoder for adding richer speech emotion information into existing speech embeddings and two multi-perspective emotion auxiliary tasks that are used only in the training stage. The implementation and the training process are detailed in Section \ref{sec:train method}.

\subsubsection{Emotion nuance module}
As aforementioned, we design an emotion nuance module in our RE-LLM, which aims to better capture the emotion information from two aspects, specifically including 1) additionally integrating emotion-related cues from speech emotion encoder, and 2) diversely considering finer dimensional emotion attributes information (valence, arousal, and dominance).

For emotion-related cues from speech, we first exploit a frozen pretrained speech emotion encoder \cite{wagner2023dawn} 
to enrich the input feature space as formulated in Eq.~\ref{eq:emo_enc}. Secondly, to better preserve emotion nuance in the embedding, we further consider finer-dimensional emotion attributes regression beyond the conventional categorical emotion classification. Hence, we integrate an auxiliary task of regressing three-dimensional emotion attributes and categorical classification at the same time. 

\subsection{Training process} \label{sec:train method}
RE-LLM training process includes two steps: pre-processing and expected behavioral alignment with emotion nuance module. The framework is shown in Figure~\ref{fig:framework}.
In the following subsections, we elaborate on each step in detail, and we define three terms used subsequently, including speech input ($S$), prompt ($P$), and text tokenizer ($Token$).

\subsubsection{Step 1: Pre-processing - expected response generation}

To capture paralinguistic features in speech and text, expected responses under specific emotions are needed. Prior work shows effective generation using emotion-conditioned prompts \cite{wang2024blspemoempatheticlargespeechlanguage} or explicit emotion labels \cite{kang2024frozen}. These responses serve as learning targets in the second alignment step.

We generate expected responses by instructing the LLM with a designed prompt to continue transcripts under the given emotion label ($<$emo$>$). The prompt is as follows:

\begin{center}
{\noindent \ttfamily \scriptsize $"$Continue the following sentence that reflects a \newline<\text{emo}> emotion tone in a coherent style:$<$transcript$>$$"$}
    
\end{center}

After collecting expected responses of each transcript input from training samples, we form it as a speech-response paired set ($\mathbb{P} = \{s_i, expect\_res_i|i=1, ..., n\}$) ready to be applied for alignment in step 2.

\subsubsection{Step 2: Expected behavioral alignment with emotion nuance module}

Expected behavioral alignment ensures an LLM responds consistently to the same input in both speech and text modalities. 

The alignment is achieved by training with our previously generated pairs set ($\mathbb{P}$). The whole process is illustrated below.
First, we integrate a pretrained speech encoder and a frozen speech emotion encoder for emotion cues. The extraction process is formulated as follows:

\begin{equation}\label{eq:emo_enc}
    Emb_{emo} = Emo\_Enc(S)
\end{equation}
\begin{equation}
    Emb_{raw} = Speech\_Enc(S;\theta_{s})
\end{equation}
where $Speech\_Enc$, $\theta_{s}$ and $Emo\_Enc$ represent a fine-tuned intrinsic speech information extractor, its corresponding parameter, and frozen pretrained speech emotion encoder, and $Emb_{emo}$ and $Emb_{raw}$ are extracted embeddings accordingly. Note that the pretrained speech emotion encoder is frozen in the entire work, so its parameter is not indicated in the formula.

Then, the modality adapter transforms concatenated speech embeddings into emotion-enriched representations for LLM response generation, operated as below:

\begin{equation}\label{eq:fuse_emb}
    Emb_{fusion} = Adapter( Emb_{raw}\oplus Emb_{emo};\theta_{adpt})
\end{equation}
\begin{equation}
    R = \text{LLM}(Token(P)\oplus Emb_{fusion}; \theta_{LM})
\end{equation}

where $Adapter$, $\theta_{adpt}$ are the modality adapter and its corresponding parameters and $Emb_{fusion}$ is its output embedding, and $R$ and $\theta_{LM}$ stand for responses and LLM parameters respectively.

Here, we also conduct a continuation writing task, ensuring the generated continuation ($R$) from speech inputs aligned with previous generated expected response ($expect\_res$). The prompt ($P$) for the continuation task is shown as below:

\begin{center}
{\noindent \ttfamily \scriptsize $"$Continue the following sentence that reflects \newline a tone in a coherent style: <speech> $"$}    
\end{center}

The response alignment is optimized by minimize KL-divergence loss \cite{wang2024blspbootstrappinglanguagespeechpretraining} formulated as below:
\vspace{-1mm}
\begin{equation}
   L_{KL}(s, expect\_res) = -\sum_{j}\log{p(y_{j} \mid \mathbf{s}, \mathbf{y}_{<j})}
   \label{eq:KL divergence}
\end{equation}
\vspace{-1mm}
where $s$ is the speech sample. $y$ is the predicted response from LLM and $j$ is the sequence index of it, and $expect\_res$ is the generated expected response conditioned on emotions from $\mathbb{P}$.

Besides, as part of our emotion nuance module, a categorical emotion classification and a dimensional regression are integrated as auxiliary tasks to provide a finer granularity through multiple perspectives in emotions. The embedding processed from Eq.~\ref{eq:fuse_emb} is a time series. We first conduct a mean pooling operation to make it an utterance-level representation before feeding into the classification and regression networks. The complete process can be formulated as below:

\begin{equation}
    Emb_{pool}= \text{Pooling}(Emb_{fusion}), Emb_{pool} \in \mathbb{R}^{B\times 1\times M}
\end{equation}
\begin{equation}
    \hat{y_{cat}} = Clf_{cat}(Emb_{pool}), \hat{y_{cat}} \in \mathbb{R}^{B\times 1 \times 4}
\end{equation}
\begin{equation}
    \hat{y_{dim}} = Reg_{dim}(Emb_{pool}), \hat{y_{dim}} \in \mathbb{R}^{B\times 1 \times 3}
\end{equation}
where $Emb_{pool}$ is the pooled utterance-level, and $Clf_{cat}$ and $Reg_{dim}$ stand for categorical emotion classifier and dimensional emotion regressor. Besides, $\hat{y_{cat}}$ is the prediction of four major categorical emotions, and $\hat{y_{dim}}$ is the regression results of dimensional emotion attributes composed of three dimensions valence, arousal and dominance for each input sample.

These auxiliary tasks refine the model's ability to capture nuanced emotions, and both cross entropy and MSE loss functions are defined below:

\begin{equation}
   \begin{cases}L_{CE} = -\sum_{i=1}^Ny_{cat, i}\log(\hat{y_{cat, i}})\\
   L_{MSE}=\frac{1}{N} \sum_{i=1}^N(y_{dim, i} - \hat{y_{dim, i}})^2\end{cases}
   \label{eq:CE loss}
\end{equation}

where $\hat{y}$ represents the emotion prediction for both perspectives and $i$ is the index of each sample, and $N$ is the total amount of training samples.
Therefore, the overall training loss is combined as below:
\begin{equation}
    L_{RE-LLM} = L_{KL} + L_{CE} + L_{MSE}
   \label{eq:all loss}
\end{equation}

\subsection{Metric of Empathy and Performance}
We evaluate empathetic quality using an automatic empathy scoring mechanism based on EPITOME \cite{sharma2020empathy}. ``Emotional Reaction (ER)" score assesses response to emotion, while ``Exploration (Ex)" score measures questioning depth, both scored from 0 (least) to 2 (most). Scores are computed using a RoBERTa model\footnote{https://github.com/behavioral-data/Empathy-Mental-Health}. We report averaged scores with standard deviations. To assess whether RE-LLM significantly outperforms the fine-tuned speech-LLM, we conduct Wilcoxon signed-rank tests on the paired ordinal scores and report its p-value. Results are shown in Table \ref{tab:result}.

For evaluation of emotion classification accuracy, we present unweighted accuracy (UA) for categorical emotion classification assessed using the prompt as the model receives the speech:

{\noindent \ttfamily \scriptsize $"$Please identify the emotion tone of the sentence provided below. Select from the following options: neutral, sad, angry, happy, or surprise.

\noindent Sentence: <speech>$"$}

\begin{table*}[ht]
\centering
\scriptsize
\begin{tabular}{l@{\hspace{2pt}}lc@{\hspace{5pt}}c@{\hspace{5pt}}cc@{\hspace{5pt}}c@{\hspace{5pt}}cc@{\hspace{5pt}}c}
\toprule
\textbf{Metric} & \textbf{Dataset} & 
\multicolumn{3}{c}{\textit{Text-LLM}} & 
\multicolumn{3}{c}{\textit{Speech-LLM}} & 
\multicolumn{2}{c}{\textit{Ablation}} \\
\cmidrule(lr){3-5} \cmidrule(lr){6-8} \cmidrule(lr){9-10}
& & 
\makecell{1.\\Text+LLM} & 
\makecell{2.\\Whisper\\+LLM} & 
\makecell{3.\\Text+LLM\\+emo\_label} & 
\makecell{4.\\BLSP-Emo\\(w/o ft.)} & 
\makecell{5.\\BLSP-Emo\\(w/ ft.)} & 
\makecell{6.\\RE-LLM} & 
\makecell{7.\\RE-LLM\\(w/o dim-emo. aux)} & 
\makecell{8.\\RE-LLM\\(w/o enc.)} \\
\midrule

\multirow{3}{*}{\makecell{Emotional\\Reaction}} 
 & IEM     & 1.782(.442) & 1.723(.489) & 1.820(.430) & 1.780(.454) & 1.771(.472) & 1.805(.419)\textsubscript{p=.025}$^{*}$& \textbf{1.856(.373)} & 1.737(.495) \\
 & ESD     & 1.609(.639)        & 1.681(.518)        & 1.730(.519)        & 1.740(.543) & 1.730(.541) & 1.847(.438)\textsubscript{p=.000}$^{***}$& \textbf{1.851(.430)}         & 1.750(.535)        \\
 & MSP-P   & 1.704(.486)        & 1.628(.548)        & 1.740(.476)        & 1.775(.483) & 1.576(.606) & 1.852(.387)\textsubscript{p=.000}$^{***}$& \textbf{1.889(.355)}         & 1.556(.387)        \\
\midrule

\multirow{3}{*}{Exploration} 
 & IEM     & 0.607(.849)        & 0.602(.850)        & 0.885(.915)        & 0.791(.948) &  0.780(.946) & \textbf{0.822(.953)\textsubscript{p=.112}}& 0.763(.944)        & 0.808(.950)        \\
 & ESD     & 0.504(.834)        & 0.592(.829)        & 1.136(.971)        & 0.820(.956) & 1.098(.968) & \textbf{1.206(.954)\textsubscript{p=.000}$^{***}$} & 1.029(.959)        & 1.114(.969)        \\
 & MSP-P   & 0.397(.770)        & 0.420(.795)        & 0.577(.879)        & 0.467(.829) & 0.521(.848) & 0.639(.898)\textsubscript{p=.005}$^{**}$& \textbf{0.739(.944)} & 0.546(.898)        \\
\midrule

\multirow{3}{*}{UA (\%)} 
 & IEM     & --           & --           & --           & 0.712        & 0.728        & \textbf{0.766}                & 0.745        & 0.729        \\
 & ESD     & --           & --           & --           & 0.960        & 0.968        & 0.983                        & \textbf{0.989} & 0.968        \\
 & MSP-P   & --           & --           & --           & 0.574        & \textbf{0.664} & 0.643                      & \textbf{0.664} & 0.661        \\
\bottomrule
\end{tabular}%
\vspace{1mm}
\caption{Empathetic performance results. Each metric (Emotional Reaction, Exploration, UA) is evaluated across datasets (IEM, ESD, MSP-P) and different model variants. Values are means (standard deviation), and $^{*}p<0.05$, $^{**}p<0.01$, $^{***}p<0.001$.}
\label{tab:result}
\vspace{-8mm}
\end{table*}

\section{Experimental Setup and Results}
\subsection{Experimental Setup}
In this work, for a fair comparison, we follow exactly the same setting in BLSP-Emo \cite{wang2024blspemoempatheticlargespeechlanguage}, where Whisper-large-v2 \cite{radford2023robust} is deployed as the speech encoder and Qwen-7B-Chat \cite{bai2023qwentechnicalreport} as the LLM. The modality adapter includes three 1D convolution layers with a bottleneck layer (512 hidden units), reducing speech feature length by a factor of 8 with stride 2, kernel size 5, and padding 2. 

In the emotion nuance module, we use a widely-adopted pretrained wav2vec 2.0  model\footnote{https://github.com/audeering/w2v2-how-to}\cite{wagner2023dawn} to extract the emotion embedding from the pre-final layer. These are concatenated with Whisper large-v2 speech embeddings at each time step for a richer emotional information.

\subsection{Dataset}
\begin{enumerate}
    \item IEMOCAP(IEM): The IEMOCAP dataset \cite{busso2008iemocap} is a benchmark for SER, containing 12 hours of dyadic speech from 10 speakers across five sessions with 10,039 utterances. Following common settings, we use 5,531 utterances covering four major emotions—neutral, happy (excited), angry, and sad—along with their valence-arousal-dominance labels. For fair comparison, we follow the same setup as BLSP-Emo \cite{wang2024blspemoempatheticlargespeechlanguage}, using utterances from sessions 1 to 4 for training and reserving session 5 exclusively for testing. 
    \item Emotional Speech Dataset (ESD): The ESD corpus, a benchmark for SER, contains 350 parallel utterances from 20 speakers (10 English, 10 Chinese) across five emotions. We use only the English data with four major emotions—neutral, happy, angry, and sad. Since dimensional labels are absent, we generate pseudo labels using the same pretrained model \cite{wagner2023dawn} for further experiments. As for data splitting, the 350 repeated utterances are divided into training and testing sets using a 7:3 ratio, ensuring a balanced emotion distribution.
    \item MSP-PODCAST(MSP-P): The MSP-PODCAST dataset\cite{8003425} is a large-scale corpus for speech emotion recognition, consisting of speech segments collected from online podcasts "in the wild". To ensure a fair comparison, we randomly sample the same number of utterances 4,290 for training and 1,245 for testing as IEMOCAP from the original MSP-PODCAST training (57,211) and testing (31,139) sets. The dataset provides both dimensional and categorical emotion annotations. Following common practice, we adopt the training set for model training and use the testing set for evaluation.
\end{enumerate}

\subsection{Baseline}

To evaluate the effectiveness of our proposed RE-LLM in the additional speech modality and the emotion nuance module, we present the following baseline models for a fair comparison, including the SOTA empathetic BLSP-Emo and its variations. All baseline models are described below:
\begin{itemize}
    \item \textbf{Text + LLM}: Ground truth transcripts as inputs to the LLM, 
    \item \textbf{Whisper + LLM}: Speech is transcribed using Whisper large-v2 \cite{radford2023robust} and directly feed forward as inputs to the LLM.
    \item \textbf{Text + LLM + emo\_label}: Categorical emotion labels (e.g., $<$sad$>$) are prepended to the transcript as inputs to the LLM.
    \item \textbf{BLSP-Emo (w/ ft., and w/o ft.)}: We use the released SOTA empathetic BLSP-Emo model to inference the response and compare with our results in both fine-tuned version and frozen version. Note that fine-tuned version is presented to avoid domain mismatch issue of the released model per se.
\end{itemize}
\vspace{-1mm}
\subsection{Results and Analysis}
To evaluate empathetic efficacy, we provide both quantitative and qualitative analyses. The comparison focuses on 1) the impact of the added speech modality and 2) the refinement by the emotion nuance module. Additionally, an ablation study with SER UA verifies the contribution of auxiliary tasks. Results are shown in Table~\ref{tab:result}.

\subsubsection{Empathetic effect wrt. speech modality} \label{sec:diffds}

To evaluate the impact of paralinguistic information, we compared text-only and text-speech models. While Model 1 (Text + LLM) and Model 2 (Whisper + LLM) process only linguistic input, Model 4 (BLSP-Emo) and Model 6 (RE-LLM) integrate speech with LLM.  

From Table~\ref{tab:result}, Model 4 and Model 6 achieve similar ER scores to Model 1 on IEMOCAP. However, on the ESD dataset, they show significant relative gains of 8.14\% and 14.79\% over Model 1 and of 3.50\% and 9.87\% over Model 2. Similarly, on the MSP-PODCAST dataset, relative improvements of 4.16\% and 8.92\% over Model 1, and 9.02\% and 13.75\% over Model 2 are observed, highlighting the benefits of paralinguistic cues.

Notably, Ex scores achieve a relative improvement of 30.31\% and 35.42\% on IEMOCAP, of 17.63\% and 60.95\% on MSP-PODCAST and surprisingly of 62.69\% and 139.28\% on ESD compared to Model 1, with similar gains over Model 2. These findings indicate that integrating linguistic and paralinguistic information not only enhances emotional responses but also improves exploration in question-asking, leading to more empathetic interactions.
\vspace{-1mm}
\subsubsection{Empathetic effect wrt. emotion nuances}
We compared three settings: Model 3 (Text + LLM + emo-label), Model 5 (fine-tuned BLSP-Emo), and Model 6 (RE-LLM) to assess the impact of emotion type and dimensional expression on empathetic responses. RE-LLM outperforms all baselines, achieving the highest ER (1.847) and Ex (1.206) scores on ESD with a relative improvement of 6.76\% and 9.83\% respectively compared to Model 5. The improvements made by RE-LLM in both ER and Ex scores over Model 5 are statistically significant almost across three datasets. Similarly, RE-LLM shows consistent improvements on MSP-PODCAST. However, its performance on IEMOCAP is comparable, suggesting limitations in handling highly dramatic emotional interactions with clear cues. These results indicate that RE-LLM, integrating a speech-LLM with an emotion nuance module, generates more effective and comprehensive empathetic responses than standard speech-LLMs. Its superior performance on ESD and MSP-PODCAST suggests better adaptability to general scenarios, as ESD especially conveys emotion more on speech tone than scripted linguistic content, unlike IEMOCAP. We even examine an extreme case in Model 3, where the ground-truth emotion label is directly provided to the LLM. Model 5 performs worse than Model 3, suggesting that standard speech-based LLMs may not reliably extract emotional cues for generating empathetic responses. In contrast, our RE-LLM achieves improvements in both ER and Ex scores over Model 3 by 6.76\% and 6.16\% on ESD and by 6.43\% and 9.18\% on MSP-PODCAST while maintaining comparable performance on IEMOCAP. This demonstrates that RE-LLM can effectively access richer emotional information, leading to more empathetic response generation.

To further verify the contribution of the emotion nuance module, we present the performance on the SER task as an indicator. For speech-based LLMs, RE-LLM achieved the highest UA (76.6\%, 98.3\%) on IEMOCAP and ESD, respectively. On MSP-PODCAST, RE-LLM achieves a UA comparable to the fine-tuned speech-LLM and outperforms the pre-trained baseline by 6.9\%. This shows that incorporating an emotion nuance module and concatenating speech emotion embedding enhances emotion recognition in speech, which in turn contributes to the generation of empathetic responses.

\begin{table*}[!t]
    \centering
    \scriptsize
    
    \begin{tabular}{!{\vrule width 1.2pt}m{1.3cm}|m{13cm}|p{0.15cm}|p{0.15cm}!{\vrule width 1.2pt}}
    \Xhline{1.2pt}
        IEM&  transcript:“What do you mean me above all look at you you’re shaking”&&\\\hline
 Model & response&Ex&ER\\\hline
        BLSP-Emo\newline(w/o ft) & {\textbf{I'm sorry to hear that you're feeling angry.} It's understandable to feel that way when someone is shaking. It can be a sign of fear or anxiety and it can be difficult to know how to respond. If you're comfortable sharing more I'm here to listen and try to help if I can.}  &0&2\\ \hline
        Text + LLM\newline+emo\_label & {\textbf{I'm sorry if I have caused you to feel angry.} I didn't mean to upset you. {\color{red}Can you please tell me what I said or did that made you feel this way?} I'm here to help and understand your perspective.}  &2 &2\\ \hline
        RE-LLM & {\textbf{I'm sorry to hear that you're feeling angry.} {\color{red}Can you please tell me more about what's been bothering you so that I can try to understand and help address your concerns?} It's important to communicate openly and honestly in order to resolve any issues that may be causing you distress.}  &2 &2\\
        \Xhline{1.2pt}
        \noalign{\vskip 1mm}
        \Xhline{1.2pt}
 ESD& 
transcript: “Why should I purchase my own?”& &\\\hline
 Model & response& Ex&ER\\\hline
 BLSP-Emo\newline(w/o ft) & {\textbf{I'm sorry to hear that you're feeling sad about the prospect of purchasing your own item.}} {\color{red}Can you please provide more context about what you're considering buying and why you're feeling sad about it?} I'm here to listen and try to help if I can.& 2&2
\\\hline
 
Text + LLM\newline+emo\_label & {\textbf{I'm sorry to hear that you're feeling sad about the prospect of purchasing your own item.}} {\color{red}Can you please provide more context so I can better understand what you're referring to and offer more specific advice?}& 2 &2
\\\hline
 RE-LLM & \textbf{I'm sorry to hear that you're feeling sad}  about having to do something that you don't want to do. It can be difficult to feel forced into actions that we don't want to take and it's understandable that you might feel sad about it. However it's important to remember that sometimes we have to do things that we don't want to do in order to take care of ourselves or others. It's possible that this particular task is important for your own well-being or the well-being of someone else and that by doing it you'll be able to move forward and feel better in the long run If you're feeling really sad about having to do this task it might be helpful to take some time to process your emotions and think about why you're feeling this way.......   & 0&2\\
\Xhline{1.2pt}
\noalign{\vskip 1mm}
\Xhline{1.2pt}
MSP-P& transcript: “but i feel like...but the spring is...particularly dense.dense for us. it's just rough for us because there's... all of our” & &
\\\hline
 Model & response& Ex&ER\\\hline
 BLSP-Emo\newline(w/o ft) & {\textbf{I'm sorry to hear that you're feeling sad.}} It's understandable to feel overwhelmed when things feel particularly dense and difficult. It's important to take care of yourself during these times and seek support from loved ones or a professional if needed. Remember that you are not alone and that things will get better. & 0&2
\\\hline
 
Text + LLM\newline+emo\_label & {\textbf{I'm sorry to hear that you're feeling sad about the dense spring.}} It can be tough when the weather doesn't cooperate with our plans or expectations. {\color{red}Is there anything specific that's causing you to feel this way?} Maybe talking about it can help you feel better.
& 2 &2
\\\hline
 RE-LLM & {\textbf{I'm sorry to hear that you're feeling sad.}} It sounds like you're going through a tough time right now. {\color{red}Is there anything specific that's causing you to feel this way?} Sometimes it can be helpful to talk about our feelings and share them with someone we trust. If you'd like I'm here to listen and try to offer some support.
 & 2 &2\\\Xhline{1.2pt}
    \end{tabular}
    \caption{Examples of responses from different settings. Ex and ER represent the Exploration score and the Emotional Reaction score correspondingly.}
    \vspace{-10mm}
    \label{tab:response case study}
\end{table*}

\subsubsection{Ablation study}

We conduct ablation studies to assess the impact of the emotion encoder and additional dimensional emotion auxiliary task in our proposed emotion nuance module. From Model 7 in Table~\ref{tab:result}, adding speech emotion embeddings without the additional dimensional emotion auxiliary task improves ER scores, aligning with Model 6 (RE-LLM). However, it creates a trade-off, slightly reducing ER scores while enhancing Ex scores except for MSP-PODCAST, indicating that the dimensional emotion auxiliary task refines emotional cues but adds complexity. 

Model 8 shows that using both auxiliary tasks without an emotion encoder boosts Ex scores but lowers ER scores. Notably, on MSP-PODCAST, this configuration results in the lowest ER across all settings and performs worse than Model 6 and Model 7 in terms of Ex ,suggesting that information from an intrinsic speech encoder alone is insufficient for such multiple perspectives emotion understanding and that simply adding emotional cues, without a multi-perspective learning framework, could impair the model’s ability to respond empathetically. These findings highlight the distinct contributions of added speech emotion embedding and auxiliary task in our proposed module.

As for more complete and balanced ability of empathy, while Model 7 achieves the highest ER among all settings, our full RE-LLM demonstrates superior overall empathetic quality compared to the categorical-only variant (e.g., ER + Ex on ESD: 3.053 vs. 2.880), making it a more effective choice.

\subsubsection{Case study}
\vspace{-1mm}
We compare responses to an angry speech prompt from IEMOCAP (transcript:``What do you mean me above all look at you you're shaking"), a sad speech prompt from ESD (transcript:``Why should I purchase my own?") and a sad speech prompt from MSP-Podcast (transcript:``but i feel like...but the spring is...particularly dense.dense for us. it's just rough for us because there's... all of our") across models. The compared models and results are shown in Table~\ref{tab:response case study}. While all models reflect the user's emotion (as highlighted in bold), only Text+LLM(+emo-label) and RE-LLM further explore the cause (mark in red) on IEMOCAP and MSP-PODCAST. This suggests that enhanced emotional information prompts LLMs to inquire about feelings, supporting our hypothesis that RE-LLM enriches emotional context like explicit emotion labels.
\vspace{-1mm}
\subsubsection{Limitation}
While our qualitative case study highlights the strength of RE-LLM in generating empathetic responses, some limitations remain. As shown in the ESD case of Table~\ref{tab:response case study}, RE-LLM achieved the same Emotional Reaction (ER) score of 2 as the other models but received an Exploration (Ex) score of 0, compared to scores of 2 for the others. Closer inspection reveals that RE-LLM produced a long, supportive message without a clear question to prompt user elaboration. In contrast, the other models included concise follow-up questions like “Can you tell me more…?”, contributing to higher Ex scores. A possible reason is that RE-LLM, due to repeated fine-tuning and behavioral alignment, tends to generate longer, monologue-style responses. Future work could explore length control mechanisms to encourage more interactive outputs.

\vspace{-3mm}
\section{Conclusions}
\vspace{-2mm}
In this work, we propose RE-LLM, a speech-based LLM that enhances empathetic AI interactions by integrating emotional embeddings and auxiliary tasks for speech emotion classification and regression. RE-LLM significantly improves Emotional Reaction and Exploration scores, surpassing previous empathetic speech-LLMs. Our results highlight the value of emotion-enhanced speech representations and leveraging dimensional emotion attributes without explicit text labels. However, RE-LLM faces challenges in highly act-out interactions like IEMOCAP. This work does not examine how the accuracy of emotion recognition impacts empathetic response quality. Moreover, our experiments focus only on single-turn utterances across three datasets. Future work will extend to multi-turn dialogues and investigate how recognition accuracy influences response quality in more detail.

\bibliographystyle{IEEEtran}
\bibliography{mybib}
\end{document}